\documentstyle[12pt]{article}                    
\newfont{\ffont}{msym10}                          
\newcommand{\beq}{\begin{equation}}               
\newcommand{\eeq}{\end{equation}}                 
\newcommand{\bqry}{\begin{eqnarray}}              
\newcommand{\eqry}{\end{eqnarray}}                
\newcommand{\bqryn}{\begin{eqnarray*}}            
\newcommand{\eqryn}{\end{eqnarray*}}              
\newcommand{\preprint}[1]{\begin{table}[t]        
            \begin{flushright}                    
            \begin{large}{#1}\end{large}          
            \end{flushright}                      
            \end{table}}                          
\newcommand{\PD}[2]                               
    {\frac{\partial^{#2}}{\partial #1^{#2}}}      
\begin{document}
\preprint{LA-UR-98-3016} 
\title{A Possible Solution to \\
the Tritium Endpoint Problem}
\author{\\ G. J. Stephenson, Jr.\thanks{E-mail:
GJS@hepv1.phys.unm.edu} \ \\  Department of Physics \&
Astronomy,\\ University of New Mexico \\ Albuquerque, NM 87131, USA
\\ and \\ T. Goldman\thanks{E-mail:  GOLDMAN@T5.LANL.GOV} \  \\Theoretical
Division, MS B283 \\  Los Alamos National Laboratory \\ Los Alamos, NM
87545, USA \\}
\maketitle
\begin{abstract}
\noindent Scalar or right-chiral interaction currents may be expected
to produce a neutrino coupled to the electron which is different from,
and perhaps even orthogonal to, that coupled to the electron by the
standard model weak interaction.  We show that, using reasonable
parameter values for such additional interactions, it is possible to
generate a spectrum which, if analyzed in the manner commonly employed
by experimental groups, produces a negative neutrino mass-squared.

\end{abstract}

\setcounter{page}{0}

\pagebreak

\section{Introduction}

Since the report of a positive result for the neutrino mass in Tritium
beta decay by the ITEP group ~\cite{Lubimov}, there have been a 
number of measurements~\cite{Zurich,LANL,LLNL,TROITSK1,
TROITSK2,MAINZ1,MAINZ2}, all of which disagree with the ITEP result and
all of which, when analyzed under the assumption of a single massive
neutrino, give the result,
generally viewed as unphysical,
that the best fit value for
$m_{\nu}^2$ is negative\footnote{For a discussion based on a tachyonic
interpretation, see~\cite{TACHYON}.}.  This has been interpreted as an
excess of counts near the end point, which could be modelled as a
``spike'' at the end point~\cite{LANL,TROITSK1}.  Furthermore, the Los
Alamos group~\cite{LANL} reported that the best fit value was dependent
on the distance from the end point taken as a cut off for including
data in the fit.  This corresponds to the dependence on the retarding
potential used in an integral experiment as reported by
Mainz~\cite{MAINZ1, MAINZ2}.

The basic idea of using the end point of a beta ray spectrum to look
for the possiblity of a neutrino mass goes back to Fermi's original
paper~\cite{Fermi}.  It was known by the late 40's~\cite{KH} that the
dependence of the spectrum near the end point on the neutrino mass is
influenced by the Lorentz nature of the Weak Interaction.  After the
introduction of parity violation, this issue was revisited, for a
single massive neutrino, by Enz~\cite{Enz}.

It is  now known that there are three Standard Model (SM) neutrinos
with masses that are small compared with half the $Z^0$ mass, and that
the neutrinos produced by the standard Weak Interaction in conjunction
with the electron, the muon and the tau correspond to three orthogonal
(at least very nearly orthogonal) combinations of these mass
eigenstates.  Therefore,
analyses of the Tritium spectrum
need to be revisited,
allowing for three separate mass eigenstates and the possibility that
other, weaker interactions may also play a role.  In a separate paper
we look at this problem in more generality~\cite{SG}.  For this Letter,
we confine our remarks to a simple argument which shows that it
is possible to generate a spectrum which would produce the ``negative
mass-squared'' effect that has been reported.

\section{The Effect of non-Standard Model Physics}

Interactions beyond the SM must be weaker, at low energies, than the
usual left-chiral SU(2) to avoid serious conflict with existing
data.  Presumably this is due to the 
boson mediating the interaction  
being much heavier than the known W's and Z, and/or the coupling
constant being smaller than that for the SM.  Absent a particular Grand
Unified Theory in which one wishes to embed the SM and the new
interactions, that is all one can say.

For low energy physics, like nuclear $\beta$-decay in general and 
Tritium $\beta$-decay in particular, such new interactions
can only appear as effective 
currents in the four fermion formulation of the theory with the usual
space time structure of  $S, P, T, V$ or $A$.  Given the dominance of the
SM, it is reasonable to recast this as $S, P, T, R$ or $L$, where $R = (V +
A)$ and $L = (V - A)$.

The effective low energy interaction Hamiltonian is
\beq
H_I = \sum_{\alpha = S,P,R,L,T} G^{\alpha}\left(
J^{\dagger}_{h\alpha}\cdot J_{l\alpha} + h.c. \right)
\eeq
where, for example,
\beq
J_{l\alpha} =\overline{\psi_l} \Gamma_\alpha \psi_{\nu}
\eeq
with $\psi_l$ representing a charged lepton, $\psi_{\nu}$ a neutral
lepton (see the discussion below for more detail), and 
a similar construction on the hadron side.  Explicitly,
\bqry
\Gamma_S &=& 1 \nonumber\\
\Gamma_P &=& \gamma^5 \nonumber\\
\Gamma_R &=& \gamma^{\mu}(1 + \gamma^5)/2
\nonumber \\
\Gamma_L &=& \gamma^{\mu}(1 - \gamma^5)/2
\nonumber \\
\Gamma_T &=& [\gamma^{\mu},\gamma^{\nu}].
\eqry

For nuclear beta decay in the SM, only $\alpha = L$
survives and $G^L$ is related to the basic parameters of
the SM by~\cite{alphaW}
\beq
G^L = V_{ud}\frac{\pi\alpha_W}{\sqrt{2}M_W^2}
\eeq
where $V_{ud}$ is the appropriate element of the
hadronic CKM matrix~\cite{CKM},  $\alpha_W$ is
the fine structure constant for the $SU(2)_W$ of the SM,
and $M_W$ is the mass of the usual $W^{\pm}$.

If there are additional bosons that couple to the left-chiral fermion
current, they can only renormalize $L$ and cannot lead to any new
structure near the end point.  For Tritium $\beta$-decay, the energy
available for the kinetic energy of the resulting particles is
approximately 18.65 keV.  Conservation of momentum and energy limits
the recoil energy of the molecular ion to a few electron volts.  We may
therefore ignore $P$ and $T$ and concentrate on the interference
between $L$ and $R$ or $S$.

In this analysis we assume three mass eigenstates $\nu_i$ with
eigenvalues $m_i$.  The charge changing current of the SM produces, in
conjuction with an electron, a linear combination of these mass
eigenstates which we call an electron anti-neutrino\footnote{Whether
the mass eigenstates are Dirac or Majorana particles, this expansion is
valid for both neutrinos and anti-neutrinos.  Where there is no
likelihood of confusion, we refer to both as neutrinos.},
\beq
\overline{\nu_e} = \sum_{i=1}^3 cos \theta_i \overline{\nu_i }\label{eq:nue}
\eeq
where the $cos \theta_i $ are the direction cosines with respect to the
three orthogonal mass eigenstates.  With this definition of
$\overline{\nu_e}$, the expected form of the differential spectrum
becomes
\beq
\frac{dN}{dE_{\beta}} = K F(Z_d ,E_{\beta}) q_{\beta} E_{\beta}
E_{\nu} \sum_{i=1}^3 cos^2 \theta_i \sqrt{E_{\nu}^2 - m_i^2}
\Theta (E_{\nu} - m_i) \label{eq:normspec}
\eeq
where $q_{\beta}$ is the electron momentum and we have explicitly
written 
\beq
q_{\nu_i} = \sqrt{E_{\nu}^2 - m_i^2} .
\eeq
The $\Theta$ functions guarantee that each mass eigenstate contributes
only above its threshhold in $E_{\nu} = E_0 - E_{\beta}$, where $E_0$
is the end point energy appropriate to a massless neutrino, and  $K$ is
a constant incorporating the SM Weak Interaction coupling constant and
the nuclear matrix elements.   $F(Z_d,E_{\beta})$ is the usual Fermi
function incorporating the electromagnetic interaction between the
outgoing $\beta$ and the daughter product denoted by $Z_d$.

Now consider the effect of another, weaker charge changing interaction,
either $R$ or $S$.  This will, {\it in general}, couple the electron to
a different linear combination of neutrino mass eigenstates
\beq
\tilde{\nu_e} = \sum_{i =1}^3 cos \tilde{\theta_i} \nu_i
\eeq
where the $cos \tilde{\theta_i}$ are the direction cosines for this new
vector.

One expects that interference effects between the new interaction and
the SM, for reactions involving charged currents and an electron, are
supressed by the overlap between $\nu_e$ and $\tilde{\nu_e}$ or by a
mass weighted overlap~\cite{SG}.  However, that assumes that all mass
eigenstates can participate.  For most measurements, $E_{\nu} \gg m_i$
and this assumption is well founded.  For $\beta$-decay end point
studies, this assumption is invalid.

The form of Eq.(\ref{eq:normspec}) rests on the fact that 
it is the mass eigenstates that propagate away from the interaction.
Therefore, the discussion of the effect on the spectrum, including
interference between $L$ from the SM and the new interaction, 
involves a separate calculation 
for each mass eigenstate.  In~\cite{SG} we discuss the 
implications of this more fully, 
including the dependence on $E_{\beta}$. In this Letter we concentrate
on the behavior within a small range below the end point (a few hundred
eV) where the variations in $E_{\beta}$ are negligible.  Given these
assumptions, Eq.(\ref{eq:normspec}) may be generalized as
\bqry
\frac{dN}{dE_{\beta}} &=& K F(Z_d ,E_{\beta}) q_{\beta} E_{\beta}
E_{\nu} \times \nonumber\\
  &    &  \sum_{i=1}^3 cos^2 \theta_i  (1+\epsilon_i )
(1 + \frac{\phi_i}{E_{\nu}})\sqrt{E_{\nu}^2 - m_i^2}
\Theta (E_{\nu} - m_i) \label{eq:newspec}
\eqry
where $\epsilon_i$ and $\phi_i$ depend on the Lorentz structure
of the new interaction.  Specifically, if
\beq
\eta = (\tilde{\alpha}/\alpha_W)(M_W^2/\tilde{M}^2)
(cos\tilde{\theta_d}/cos\theta_d)
\eeq
where, as before, $\alpha_W$ is the usual Weak Interaction fine structure
constant~\cite{alphaW}, $\tilde{\alpha}$ that for the new 
interaction,
$M_W$ is the mass of the usual $W^{\pm}$ bosons and
$\tilde{M}$ is the mass of the charged bosons mediating
the new interaction.  $cos\theta_d$  and $cos\tilde{\theta_d}$
are the analogous direction cosines in the hadronic sector
\footnote{In the hadron sector, these direction cosines are
usually expressed, for the SM, as elements of the CKM 
matrix~\cite{CKM}.}.

Then, for $R$,
\bqry
\xi_i &=& \eta (cos\tilde{\theta_i}/cos\theta_i) \nonumber\\
\phi_i &=& \frac{2m_i \xi_i }{(1 + \epsilon_i)}\left(\frac{m_e}{E_0}
\right)  \nonumber\\
\epsilon_i &=& \xi_i^2 \label{eq:R}
\eqry
while for $S$,
\bqry
\xi_i &=& \eta (cos\tilde{\theta_i}/cos\theta_i) \nonumber\\
\phi_i &=& -\frac{m_i (\xi_i +\frac{2\xi_i^2 m_e}{E_0})}{1+ \epsilon_i} \nonumber\\
\epsilon_i &=&  \xi_i \frac{m_e}{E_0}+2 \xi_i^2 \label{eq:S}
\eqry
The sign in Eq.(\ref{eq:S}) comes from performing the usual traces, but
the overall sign of the $\xi_i$ is indeterminate due to the various
direction cosines.  (There is, in fact, additional freedom due to
purely left- or right-chiral couplings of S to leptons\cite{Herczeg},
which are treated in \cite{SG}.)

\section{Effect on Differential Spectra}

Let us now assume, for simplicity, that only one mass eigenstate is
important for the fit near the end point, given the resolution of
current experiments.
Let that mass eigenvalue be denoted as $m_1$, and scale
$E_{\nu}$ and the $\phi_i$ in units of $m_1$,
\bqry
x & = & E_{\nu}/m_1 \nonumber \\
f_i & = & \phi_i /m_1
\eqry
In this case, Eq.(\ref{eq:newspec}) becomes
\beq
\frac{dN}{dE_{\beta}} = K'(1+\epsilon_1) x^2 cos^2\theta_1
(1+\frac{f_1}{x})\sqrt{1 - \frac{1}{x^2}}
\Theta(x-1)  \label{eq:simp1}
\eeq
If we were to fit this with a formula for the spectrum which is derived
under the assumption that there is only one neutrino and that it has
a mass extracted from the spectrum as $<m^2>_{fit}$,
we would fit Eq.(\ref{eq:simp1}) with the function
\beq
\frac{dN}{dE_{\beta}} = K'(1+\epsilon_1) x^2\sqrt{1-\frac{<r^2>_{fit}}{x^2}}
\label{eq:simp2}
\eeq
where $<m^2>_{fit} = m_1^2 <r^2>_{fit}$  is the extracted 
(apparent) value of 
the neutrino mass-squared. 
\newpage

Setting Eqs (\ref{eq:simp1}) and (\ref{eq:simp2}) equal, at some
particular value of x, gives an equation for $<r^2>_{fit}$,
\bqry
<r^2>_{fit} &=& x^2[1-cos^4\theta_1] \nonumber\\
                     &  & - x[2 f_1 cos^4\theta_1] \nonumber\\
                     &  & + (1-f_1^2) cos^4\theta_1 \nonumber\\
	       &  & + x^{-1} 2f_1 cos^4\theta_1 \nonumber\\
	       &  & + x^{-2} f_1^2 cos^4\theta_1
\label{eq:mvsx} 
\eqry
At $x = 1$, precisely at the end point of the physical spectrum, Eq.
(\ref{eq:mvsx}) gives the result $<r^2>_{fit} = 1$.  If the quantity
$cos^4\theta_1$ is small compared to $1$, $<r^2>_{fit}$
will grow as $x$ increases.  On the other hand, if this quantity is
nearly $1$, which would occur if $\overline{\nu_e}$ is nearly a mass
eigenstate, then the second term in Eq.(\ref{eq:mvsx}) can dominate,
leading to $<r^2>_{fit} < 0$ for some values of $x$.  The value of 
$x$ at which $<r^2>_{fit}$ again becomes positive is a sensitive
function of $f_1$ and $cos^2\theta_1$.

\section{Effect on Integral Spectra}

To obtain sufficient statistics, experiments that measure differential
spectra make a global fit to data over some range from the end point
up to some value of $E_{\nu}$, which translates, in practice, to
fitting over a range of $E_{\beta}$ down to some value.  This is done
automatically in those experiments that measure an integral spectrum.

In these cases, the fitting procedure, viewed in terms of theoretical
constructs only and ignoring 
essential experimental details, corresponds to finding the value of
$<r^2>_{fit}$ that minimizes the integral
\beq
I = \int_1^{x^c} dx x^4 (1+\epsilon_1)^2 \left[cos^2\theta_1
(1+\frac{f_1}{x})\sqrt{1-\frac{1}{x^2}} -
\sqrt{1-\frac{<r^2>_{fit}}{x^2}}
\right]^2  \label{eq:int}
\eeq

\section{A Numerical Example}

To illustrate the effects on the extraction of $<m^2>_{fit}$, we have
chosen a particular example.  This choice is motivated solely by
pedagogy.  We assume 
\bqry
f_{1} & = & 0.2/(1.01) \nonumber\\
\tilde{\theta_1} & = & - \theta_1 \nonumber\\
cos^2\theta_1 & = & 0.995
\eqry
The sign of $\tilde{\theta_1}$ plays no role here but, as shown
in~\cite{SG}, when coupled with the assumptions that only two mass
eigenstates contribute and $m_2 = 200 m_1$, it guarantees that the
interference cancels far from the end point, leaving only a small
normalization correction to the standard spectrum.

The actual effect on the spectrum is quite small. 
Nonetheless, the impact on $<m^2>_{fit}$ is marked, as can be
seen in the Figure.  To obtain that curve, we integrated the expression in
Eq.( \ref{eq:int}) to $x^c$ for a grid of $<r^2>_{fit}$, then found, by
a three point interpolation, the value that minimized $I$.  For this
example, one can clearly see the dependence of the extracted
$<m^2>_{fit}$ on the cut-off energy.  Furthermore, if one argued for
choosing a cut-off where the extracted value was most stable, one would
obtain $<m^2>_{fit} \approx -2 m_1^2$.  In spite of the integration in
Eq.(\ref{eq:int}), the sensitivity to parameter values discussed after
Eq.(\ref{eq:mvsx}) persists.  This suggests the possibility of extracting
$f_1$ and $cos^2\theta_1$ directly from a curve of $<m^2>_{fit}$
versus the cut-off energy.

\section{Conclusion}

In summary, any charge-changing interaction beyond the SM is likely to
couple the electron to a different combination of neutrino mass
eigenstates than that coupled to the electron by the SM.  This can have
the effect of {\it decreasing} expected interference signals in regimes
where all mass eigenstates participate fully.  Near the end point of a
beta ray spectrum, however, some mass eigenstates may be kinematically
excluded, restoring the visibility of interference signals.

We have recast the usual analysis in terms of the mass eigenstates to
make this possibility more transparent, and have shown that it is quite
possible to obtain values of  $<m^2>_{fit} < 0$.  We therefore
recommend that the form of Eq.(\ref{eq:simp1}) be used in the analysis
of the Tritium $\beta$-decay spectrum.

\section{Figure Caption}

\hspace*{0.18in} Figure 1.   The value of $<r^2>_{fit}$ that
minimizes the integral
expression of Eq.(\ref{eq:int}) as a function of the cut-off energy
expressed as $E_{\nu}^c/m_1$.  The parameters of the calculated
spectrum are given in the text; the value of $f_1$ has been rounded
in the label.

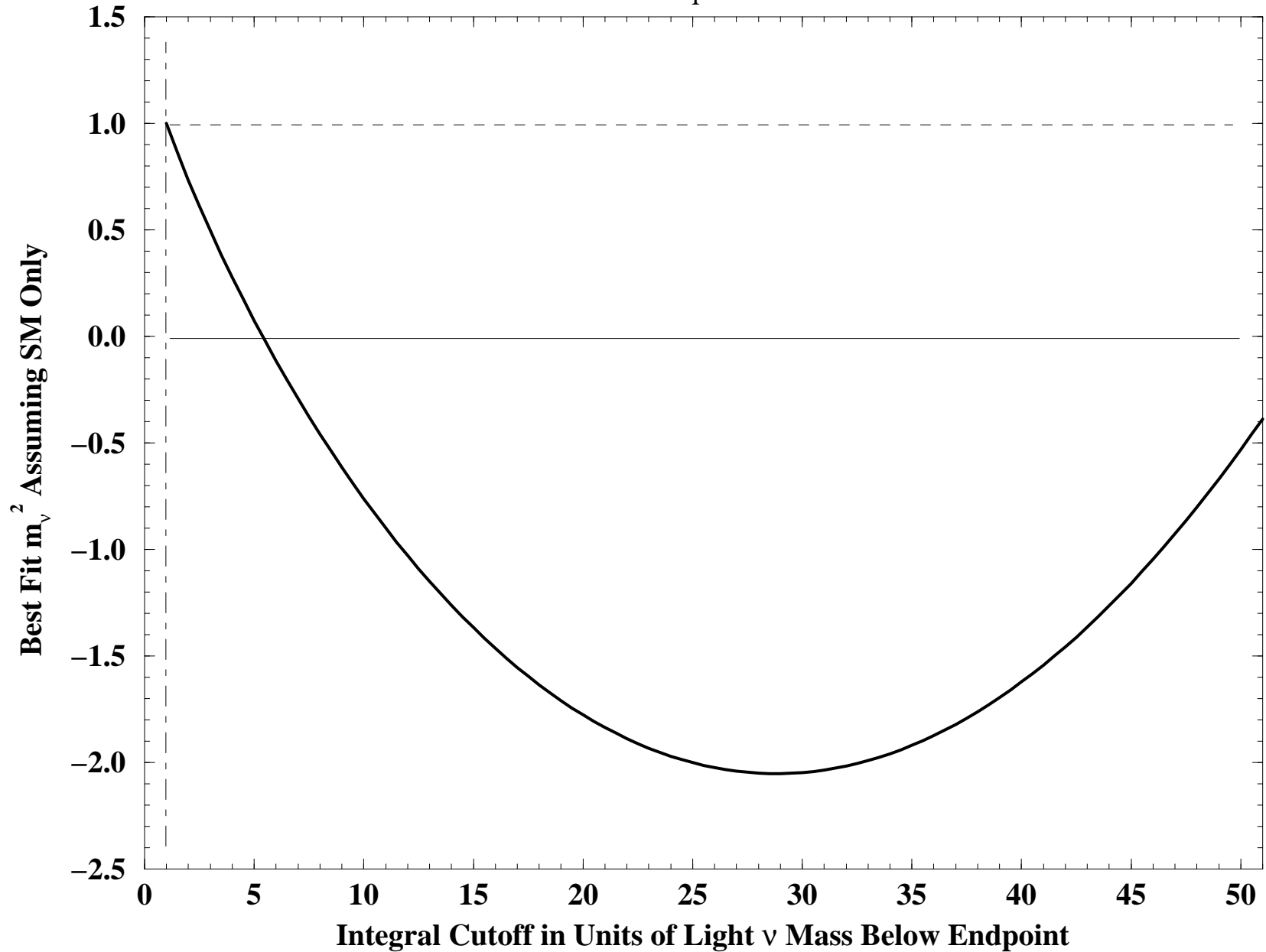
\end{document}